\documentclass[11pt,twoside]{book}
\usepackage{konkolyproc2}
\usepackage{longtable}
\usepackage{amsmath,amssymb}
\usepackage{graphicx}
\usepackage{lscape}
\usepackage{index}
\usepackage{natbib}
\usepackage{bigdelim}
\usepackage{multirow}
\usepackage{enumerate}

\makeindex

\begin{document}

\pagestyle{myheadings}
\setcounter{equation}{0}\setcounter{figure}{0}\setcounter{footnote}{0}\setcounter{section}{0}\setcounter{table}{0}\setcounter{page}{1}
\markboth{Muthsam \& Kupka}{Multidimensional Modelling of Classical Pulsating Stars}
\title{Multidimensional modelling of classical pulsating stars}
\author{Herbert J. Muthsam$^1$ \& Friedrich Kupka$^1$ }
\affil{$^1$Faculty of Mathematics, University of Vienna, Austria}

\begin{abstract}
 After an overview of general aspects of modelling the pulsation-convection interaction we present reasons why such simulations (in multidimensions) are needed but, at the same time, pose a considerable challenge. We then discuss, for several topics, what 
 insights multidimensional simulations have either already provided or can be expected to yield in the future. We finally discuss 
 properties of our ANTARES code. Many of these features  can be expected to be characteristic of other codes which may possibly be 
 applied to these physical questions in the foreseeable future. 
\end{abstract}

\section{Introduction}
Numerical modelling RR Lyr stars and Cepheids\footnote{These classical variables, basically exhibiting \textit{radial} oscillations, will be termed here as ``pulsating stars" for the sake of brevity.} has been one of the earliest successes of modern computational stellar physics. Computers becoming  available around 1960, say, allowed for the first time to build numerical models which encompass a physically  ``complete" (in some sense) description of ``real" astrophysical objects. It cannot come as a surprise that such early work centered on questions  concerning objects which are (or were considered  as) planar or spherically symmetric and therefore, in a computational sense, one-dimensional (1D) plus possibly time-dependent. This comprised issues in stellar atmospheres, stellar structure and evolution, but also protostellar collapse, supernova explosions, etc. Modelling pulsating stars fell into that category, and relatively soon after ultimate identification of the basic source of pulsation excitation even nonlinear, time dependent models became available. An overview of early  can be gained by consulting Zhevakin (1963), Christy (1966), Christy (1966a).  Nonlinear time-dependent calculations proved particularly useful in providing nontrivial properties of, for example, the light curve (departure from the sinusoidal form). That allowed stringent confrontations with and deductions from observations. In a number of cases a mismatch between observations and simulations resulted.

Today's basic source for mismatch is quite likely the pulsation-convection interaction. The crucial role of the interaction between convection and pulsation
for a number of observable properties of classical variables has been
discussed in the literature for a long time. Examples include the reviews 
of Buchler (1997, 2009) and Marconi (2009). Of particular interest
in this context is the boundary between objects exhibiting large radial 
pulsations due to the $\kappa$-mechanism on the ``hotter'' or ``blue''
side of this transition region and stable objects on the ``cooler'' or
``red'' side of this famous ``red edge'' of the pulsational instability strip.
Other, related observations include the presence of double- or even
triple mode pulsators and their location as a function of effective temperature
and luminosity or any observational equivalent and the particular shapes
(amplitudes, phase dependence) of lightcurves observed for these objects.
Since for sufficiently low effective temperature the blocking of radiative
flux by the partial ionization of hydrogen and its associated increase 
in opacity becomes very large, deep convection
zones have inevitably to form for these objects to  
take care of energy transport instead of radiation.

The ``classical" 1D simulations had by necessity to include the pulsation-convection interaction in a manner compliant with this basic computational setting. Hence, they heavily parametrized this interaction so that multidimensional simulations which \textit{in principle} are capable to fully deal with this issue are warranted. Indeed, on  the side of theoretical modelling two grand numerical challenges have been identified in the area of pulsating stars after a critique of the present hydrocodes (Buchler(2009)): \textit{Multidimensional simulation of convection in classical pulsating stars} and \textit{Modelling of nonradial pulsation} in such stars \footnote{Of course, there are unsolved problems, in particular the Blazhko effect, where even the basic physical cause is not yet known with certainty; hence it is also unknown what precise sort of computational experiments would be required to deal with it.}.

After very early work (Deupree(1977) and subsequent papers) on multidimensional modelling of convection in pulsating stars the topic has been considered anew only in the last few years. Of course, this research greatly benefits from the huge increase in computer power as well as from advances in numerical methods and codes. Still, such work amounts to a great challenge even today.

In the present paper we want to discuss first \textit{why} modelling pulsation-convection interaction in multidimensions poses so large difficulties. We will then turn towards various scientific questions and the way they have recently been investigated or which might benefit in the future with the methods available or within reasonable reach. Since the contribution of Deupree (2016) in these Proceedings covers what is being done in this group  we will concentrate on such items which are close to our own research and its possible extensions.

A discussion of the limitations of the approach is also warranted in order to draw a realistic pathway of possible future development. We therefore also give room to such considerations. From these discussions it will follow that various questions in this area cannot be directly tackled by multidimensional simulations, and some can be approached this way only at a cost which is prohibitive for everyday work. It is therefore essential that improved convection models are developed which comply with the basic setting of traditional codes (spherically symmetrical and, hence, geometrically 1D). We discuss some basic requirements such models would have to fulfill.

The numerical simulation software is equally  important as the enhanced power of computers. For that reason we give an overview of the capabilities of our ANTARES code (Muthsam, Kupka, L{\"o}w-Baselli et al. (2010), Mundprecht, Muthsam \& Kupka (2013)) as a guide what will be necessary for other 
codes which should be applied for these problems in the future. 

\section{Pulsation-convection interaction in multidimensions: the challenge}

From a numerical point of view, already the 1D (radially symmetric) simulations faced the problem of a very steep temperature gradient and  small characteristic spatial scales in the hydrogen ionization region. That led to the introduction of adaptive gridding strategies as described in papers by Castor, Davis \& Davison (1977), Dorfi \& Feuchtinger (1991), Buchler, Koll{\'a}th \& Marom  (1997).

This already explains in part why, with the exception of very early work by Deupree (1977) and some subsequent papers, modelling of pulsating stars has remained in the 1D setting for so long. Only recently has hydrodynamic modelling of pulsating variables gained impetus as is being described by Deupree's contribution to these {Proceedings} and by the present article.

This late advent of multidimensional models for pulsating stars may appear particularly conspicuous given how advanced 3D modelling of solar granulation, the issue probably best explored in this area, has been already quarter of a century ago (e.g. the review by Nordlund 
\& Stein (1990)). One reason for this different pace of progress is the presence of the small spatial scales just mentioned. This makes also modelling of even the \textit{atmospheres} only of some other types of stars a challenge. It applies, for example, to ordinary main-sequence A-type stars as extensively discussed in Kupka, Ballot \& Muthsam (2009). Lack of resolution is probably the source of  poor agreement between observed and simulated spectral line shapes in such objects  (Kochukhov, 
Freytag, Piskunov et al., (2007)).

The other major reason has to be found in the fact that in pulsating stars the timescales of convection and pulsation are comparable, largely precluding, for example, taking a fixed phase in the sense of pulsation and letting convection develop against this static background. For a fair comparison of the challenges inherent in convection modelling for solar-like and pulsating stars, respectively, it must however be kept in mind that there is more to modelling solar like stars than may be deduced from the considerations above. Modelling of solar granulation is just a part of the task, made relatively easy by the fact that the solar atmosphere is not heavily influenced by what is going on in deeper layers (setting sunspots etc. aside). Modelling \textit{all} of the solar convection zone is a challenge however due to the huge disparity in time scales (minutes or even a few seconds at the top, and hundred thousands of years at the bottom). As a consequence, two largely disjoint research communities have developed in this area, with focus either on the outer layers and on the interior. For reviews on the state of art in these fields see Miesch, Matthaeus, Brandenburg et al. (2015). Because of the long time-scales even models of the bulk zone of solar convection (omitting the surface layers) one has to modify physical parameters in order to get manageable relaxation times near the bottom. In pulsating stars this problem may either not exist as in  Mundprecht, Muthsam \& Kupka (2015) or render multidimensional simulations unfeasible as in 
Geroux \& Deupree (2013) (even if probably in a less fundamental sense than in the solar case), depending on the model under investigation.

To sum up, modelling \textit{all} of the solar convection zone (setting aside 
magnetic effects) has to consider 
\begin{enumerate}[i)]
\item{huge differences in spatial and time-scales at top and bottom (with top, i.e. granulation
readily amenable to 3D modelling)}
\item{extremely large time scales also at the bottom which request changing parameters 
to end up with a computable problem even if excluding the top of the convection zone.}
\end{enumerate}

As a consequence, there exists no simulation of the solar convection zone which 
represents both granulation and the bulk of the zone. -- In pulsating stars, atmospheres may put a problem 
even if they were treated in 
isolation, but, other than in the sun, 
they cannot be essentially decoupled from the deeper layers. So, one has to model the whole 
star. Solar problem ii) may or may not exist in such stars, depending on the star under 
consideration. 

Given these facts we address now achievements of  multidimensional models of pulsating 
stars  and we discuss specific points which could be clarified in the 
foreseeable future by such methods. Such items concern specific questions either immediately,
or they might help improving traditional 1D modelling.

\section{Multidimensional models: present and future insights}

\subsection{Testing time dependent convection models (TDC's)}

In  Mundprecht, Muthsam \& Kupka (2015) the He{\sc{ii}} convection zone has been investigated in a 
2D Cepheid model. Besides studying mechanical driving and damping such simulations allow to 
test traditional convection models such as those of Kuhfu{\ss} (1986) and Stellingwerf (1982). These time dependent convection models (TDC's) allow to stay with the 1D nature of classical simulations of pulsating stars in the numerical sense. A calibration factor
$\alpha_c$ which gives the ratio between the 1D model flux and the physical flux varies by a 
factor of $\sim 2$ during a pulsation. Even more worrysome is that $\alpha_c$, evaluated either in the
convection or overshoot zone, varies by a factor of up to 10, exact numbers depending on details. This results 
in much too strong an overshoot zone in the 1D models and casts doubts on results which may be sensitive on convective properties.
 Therefore, further development of TDC's is of paramount importance. We turn to some details now.

\subsection{Further development of TDC's}

Rather varied
 models (compliant with the 1D setting) have been proposed to describe how to compute the 
convective (non-radiative) energy transport in this case and how 
convection influences the large scale, radial pulsations in these objects.
Buchler \& Koll\'ath (2000) give a summary of the models most widely
used in this context, in particular those of  Kuhfu{\ss} and Stellingwerf just mentioned. 

In one way or another each of these models provide a prescription to
compute ensemble averages of the dynamical variables and quantities
of interest which are needed to construct a closed system of predictive
equations. The problem with each of these models is that the involve
a lot of \textit{ad hoc} reasoning which in turn has been driven by keeping the
models as simple as possible. Thus, they usually involve only one time
dependent equation joined by a number of algebraic relations and a host
of modelling parameters (up to eight) which cannot be determined from 
within the model and for which there is no reason to assume that they
are universal in the sense that once they have been calibrated for
one star by some kind of measurements they may also be used to model 
other stars without invoking further changes to those modelling parameters.

Rather than attempting to fit some set of observations with sufficiently
many degrees of freedom due to the number of available model parameters,
one might ``turn the table around'' and ask instead: what is the minimum
complexity a model has to have in order to describe the physics of
convection-pulsation interaction properly? With numerical simulations at 
hands this is now a question we can hope to provide some answers for.
In Mundprecht, Muthsam \& Kupka (2015) a model of the convective flux --- as discussed
by Buchler \& Koll\'ath (2000) and as used by many investigators who
use either the convection model of Stellingwerf (1982) or that one by
Kuhfu{\ss} (1986) in their studies of classical variables --- was analyzed
in detail. It turned out that a proper reproduction of the convective flux 
computed from the numerical simulation requires a model to 
a) account for the efficiency of convection as a function of distance
to the boundary of the convection zone, b) account for this efficiency
to behave completely differently in stable and unstable layers, 
c) consider the dependence of this efficiency as a function of
pulsation phase, d) consider that the ratio of kinetic energy stored
in horizontal motions versus that one stored in vertical motions is different
inside the convection zone (where vertical flows dominate) and outside
of it (where horizontal ones dominate), and e) the convective flow
is subject to radiative losses. While the latter is typically accounted for,
the success in doing so for the others is either limited or they are
ignored altogether. 

Reynolds stress models such as those proposed by Canuto (1993, 1997, 1999),
Canuto \& Dubovikov (1998), or more recently in Canuto (2011), or alternatively
Xiong \& Deng (1997), provide at least the formal framework to take these properties 
into account. With separate dynamical equations for total and vertical kinetic energy, 
temperature fluctuations and velocity-temperature cross-correlation (convective flux), 
and optionally also the dissipation rate of kinetic energy and a coupling to mean 
structure equations which may even feature a non-zero mean radial velocity, 
the dependencies of the convective flux observed in the numerical simulations
could be obtained from within the model itself. Whether this is indeed the
case or how these much more complex models would have to be adjusted
to succeed remains an open question. Multidimensional simulations would be an 
important tool to guide such model development and to ultimately provide test cases.

It goes without saying that such tests would be equally important for physically and 
algorithmically much simpler models than those of Reynolds stress type such as, for example, 
presented by 
Pasetto, Chiosi, Cropper \& al. (2014) where, contrarily to what holds true 
 in Reynolds stress models, effects of compressibility,
turbulence and, in its present form, also overshooting are not included quite from the outset.

\subsection{The atmospheres}

Spectroscopy clearly demonstrates the presence of phase-varying turbulence or radial shocks 
in many pulsating stars, see e.g. Preston (2011) for RR Lyr stars. Our simulations 
for a Cepheid model have shown strong, phase-varying nonradial shocks (Mundprecht, Muthsam \& Kupka (2013)). As a consequence, adding this to the peculiarities of the atmospheres we have 
mentioned previously the atmospheres are a challenging and rewarding object for numerical investigation in multidimensions. They require high resolution so that even 2D simulations are 
expensive. 3D simulations would demand for extremely massive computational resources at least for 
the objects where we presently have an estimate of the requirements. Since convection in the atmosophere can easily be 
much more violent than convection in the He{\sc{II}} zone with strong shocks development of 
faithful TDC's for that purpose may amount to a task even tougher  than holds true for the bulk zone.

\subsubsection{Determination of basic stellar parameters.}

Multidimensional atmospheric models would obviously bring about benefits in conjunction with the determination of the star's basic parameters (${T}_{\rm eff}, \log g, \mbox{abundances,\dots})$. Note that with very few exceptions all work in that direction has been based on static, planar model atmospheres if not even resorting to curve-of-growth methodology. 

This must provoke concerns already on general grounds given that the interaction of pulsation with atmospheric structure is known to be strong in a number of cases (see e.g. Preston (2011)). There are other hints which point to major effects of pulsation on atmospheric structure and, hence, 
derived parameters. So, for a Cepheid which at the same time happens to be a member of an eclipsing binary it has been possible to determine the limb darkening coefficient. In that analysis  
Pilecki, Graczyk, Pietrzy{\'n}ski et al. (2013) found a much lower limb darkening coefficient 
for the Cepheid component than would result from a static atmosphere with the same basic 
parameters. In their paper, Barcza \& Benk{\H o} (2014) note difficulties when applying their method to 
determine fundamental parameters, based among others on broad-band colours from ATLAS atmospheres, in particular for one star (DH Peg). 

\subsubsection{Broad-band investigations.} The remarks just made prompt of course the need for assessing the difference between 
broad-band colours obtained from ATLAS-type and fully dynamic atmospheres. A similar need for studies of the atmosphere arises for the limb darkening coefficient and 
appropriate determination of the p-factors critical for distance determination of pulsating stars via the Baade-Wesselink method. This 
factor depends on the physics of the star's atmosphere in a complicated way (e.g. Nardetto, Storm, Gieren et al. (2014)). 

A special way of interaction of the hydrogen ionization front with the photosphere of a pulsating star has implications 
for the relations between period, colour and amplitude, see Kanbur \& Ngeow  (2006). Due to the inhomogenous nature of the 
atmospheres of such stars a multidimensional study is warranted.

\subsubsection{Nonradial pulsations?} We have observed a tantalizing phenomenon when calculating an annulus containing the 
He{\sc{II}} convection zone of a Cepheid (no atmosphere) and regions below to fully allow for overshoot (Kupka, Mundprecht \& Muthsam (2014)). This models was 
developed for 20 pulsations of full amplitude. In that model the convective cells flock together in certain regions as time advances, leaving other 
places at the same radial distance essentially void. In otherwise similar models with sectors of a ``normal" opening angle ($10^{\circ•}$) 
we could not observe that phenomenon. One wonders whether in this way nonradial modes can be excited. If so, it would obviously be difficult 
to model such a phenomenon in the TDC setting since it is highly questionable whether one can properly represent such a genuinely geometrical and collective 
behaviour of convection cells in a 1D model. It would be of importance to corroborate such an effect by running several multidimensional models.

\section{Multidimensional models: code requirements}

As it seems, the classical hydrocodes for radial stellar pulsation have been developed with precisely the goal of 
modelling stellar pulsation in mind. In contrast, our ANTARES code has been developed differently, as a  
general tool for research in stellar physics or astrophysics with special features added to make is also useful for research in 
stellar pulsation. Since even a multidimensional code streamlined towards that goal will be quite large codes used in the future 
may be derived from  (or later on extended to) a general, multipurpose package. Properties of the ANTARES code which we deem to be essential 
for future codes in this area include
\begin{itemize}
\item{time-dependent radiation-hydrodynamics in 1D, 2D, 3D}
\item{radiative transfer (grey or nongrey by opacity binning), diffusion approximation in the interior}
\item{high resolution numerical schemes}
\item{polar coordinates\footnote{If, in 3D, a whole spherical shell should be modelled polar coordinates are unsuited because of the 
convergence of longitude circles (at constant radius) towards the polar axis. Placing the star in a box may be feasible but leads to 
extremely serious problems with resolution (or demands very advanced grid-refinement strategies) in the atmosphere for which the box coordinates are ill suited. One way out of that difficulty could be the development and application of curved grid techniques with irregular meshes; see e.g.  Grimm-Strele,  Kupka \& Muthsam (2014).}}
\item{coordinates moving radially, adapting to pulsation}
\item{grid refinement, at least for logically rectangular patches}
\item{highly paralellizable (in ANTARES based on MPI plus OpenMP for parallelization \textit{within} multicore nodes) }

\end{itemize}

\section{Conclusions} 

We have given a comparison of the difficulties in modelling radial pulsations of convective stars with modelling the solar 
convection zone, the most advanced topic in stellar convection studies. Multidimensional models show a very substantial weakness 
of convection models traditionally applied in 1D. Multidimensional models are capable to tackle a number of specific problems 
directly. For other questions, they may aid development and testing of TDC's. High quality numerics and a 
number of advanced code features are necessary for many problems in that area.

\textbf{Acknowledgements} We are thankful to Eva Mundprecht for long lasting cooperation. This work has benefited directly or indirectly from several Austrian Science Foundation ASF projects (P18224, P21742, P25229).


\end{document}